\documentclass[aps,prb,twocolumn,showpacs]{revtex4}

\usepackage{graphicx}
\usepackage{amsmath}
\usepackage{amssymb}

\begin{document}

\title{Correlation between charge order and second-neighbor hopping in cuprate superconductors}

\author{Huaisong Zhao}

\affiliation{College of Physics, Qingdao University, Qingdao 266071, China}

\author{Yingping Mou and Shiping Feng}
\email{spfeng@bnu.edu.cn}

\affiliation{Department of Physics, Beijing Normal University, Beijing 100875, China}

\begin{abstract}
The correlation between the charge-order wave vector $Q_{\rm CD}$ and second-neighbor hopping $t'$ in cuprate superconductors is studied based on the $t$-$t'$-$J$ model. It is shown that the magnitude of the charge-order wave vector $Q_{\rm CD}$ increases with the increase of $t'$, and then the experimentally observed differences of the magnitudes of the charge-order wave vector $Q_{\rm CD}$ among the different families of cuprate superconductors at the same doping concentration can be attributed to the different values of $t'$.
\end{abstract}


\maketitle

The charge-order correlation, as a competing phenomenon to superconductivity, is a common feature of cuprate superconductors \cite{Comin15,Comin14,Wu11,Chang12,Ghiringhelli12,Neto14,Campi15,Comin15a,Hashimoto15}. In particular, the electron Fermi surface (EFS) measurements in the angle-resolved photoemission spectroscopy (ARPES) experiments demonstrated a close connection between the charge-order wave vector $Q_{\rm CD}$ and the wave vector connecting the tips of the Fermi arcs \cite{Comin15,Comin14,Neto14}, which in this case coincide with the hot spots on EFS, providing an evidence that the charge-order correlation is a natural consequence of the EFS instability. However, the magnitudes of the charge-order wave vector $Q_{\rm CD}$ are different for the different families of cuprate superconductors at the same doping concentration \cite{Comin15,Comin14,Wu11,Chang12,Ghiringhelli12,Neto14,Campi15,Comin15a,Hashimoto15}. In this case, a systematic investigation of the different magnitudes of the charge-order wave vectors among the different families of cuprate superconductors is useful for the understanding of the physical origin of the charge-order formation. Very soon after the discovery of superconductivity in cuprate superconductors, Anderson \cite{Anderson87} argued that the essential physics of cuprate superconductors is contained in the $t$-$J$ model on a square lattice, with $t$ and $J$ that are the nearest-neighbor hopping and nearest-neighbor spin-spin antiferromagnetic exchange, respectively. However, the experimental data detected from various techniques have introduced important constraints on the microscopic model \cite{Damascelli03,Campuzano04,Wells95,Kim98,Tanaka04}. In particular, the early ARPES experiments \cite{Wells95,Kim98,Tanaka04} indicated that the electronic structure and the related overall EFS in cuprate superconductors can be properly described only by generalizing the $t$-$J$ model to include the second- and third-nearest neighbor hopping terms $t'$ and $t''$. Furthermore, the experimental analysis \cite{Tanaka04} also showed that the superconducting transition temperature $T_{\rm c}$ in the different families of cuprate superconductors is strongly correlated with $t'$. However, it has been shown experimentally that the values of $t$ and $J$ are almost common for the different families of cuprate superconductors, and only the values of $t'$ and $t''$ are different \cite{Damascelli03,Campuzano04,Wells95,Kim98,Tanaka04}. In this case, a question is raised: is there a correlation between the experimentally observed differences of the magnitudes of the charge-order wave vector $Q_{\rm CD}$ among the different families of cuprate superconductors at the same doping concentration and the second-nearest neighbor hopping $t'$.

\begin{figure*}[t!]
\centering
\includegraphics[scale=1.0]{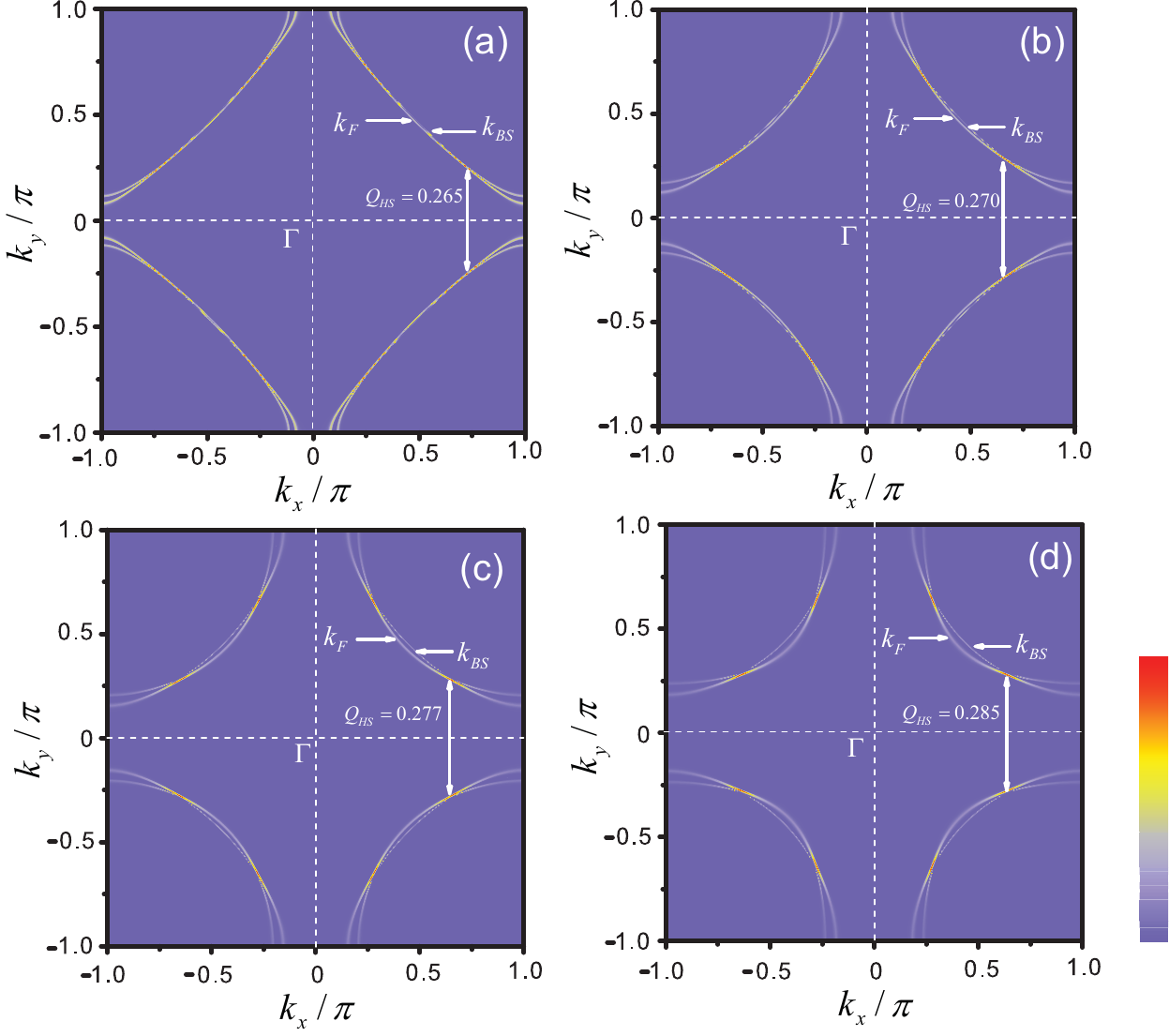}
\caption{(Color online) The spectral intensity maps $A({\bf k}, 0)$ at the Fermi energy with doping concentration $\delta=0.12$ for (a) $t'/t=0.1$, (b) $t'/t=0.2$, (c) $t'/t=0.3$ and (d) $t'/t=0.4$ with $t/J=2.5$ in $T=0.002J$. \label{spectrum}}
\end{figure*}

In the recent studies, the nature of the charge-order correlation in cuprate superconductors and its evolution with doping have been discussed based on the $t$-$t'$-$J$ model in the charge-spin separation fermion-spin representation \cite{Mou17,Feng16,Zhao16}, and the results show that the charge-order state is driven by the EFS instability, with a characteristic wave vector corresponding to the hot spots of the Fermi arcs. Although the Fermi arc increases its length as a function of doping, the charge-order wave vector reduces almost linearity with the increase of doping \cite{Feng16}. Moreover, this charge order generates a coexistence of the Fermi arcs and Fermi pockets \cite{Zhao17}. In this paper, we discuss the correlation between the charge-order wave vector $Q_{\rm CD}$ and second-neighbor hopping $t'$ in cuprate superconductors along with this line. Our result shows that the magnitude of the charge-order wave vector $Q_{\rm CD}$ increases with the increase of $t'$, and then the experimentally observed differences of the magnitudes of the charge-order wave vector $Q_{\rm CD}$ among the different families of cuprate superconductors at the same doping concentration \cite{Comin15,Comin14,Wu11,Chang12,Ghiringhelli12,Neto14,Campi15,Comin15a,Hashimoto15} can be attributed to the different values of $t'$. Our result also shows that the EFS shape itself is influenced by the second-neighbor hopping $t'$.

The charge-order correlation features a tendency toward a periodic self-organization of the charge degrees of freedom in the system \cite{Comin15}, and such a physical property should be reflected in the low-energy electronic structure. To discuss the correlation between the charge-order wave vector $Q_{\rm CD}$ and the second-nearest neighbor hopping $t'$, we first need to understand the nature of EFS, which is determined by the poles of the electron Green's function. Within the $t$-$t'$-$J$ model in the charge-spin separation fermion-spin representation, the electron Green's function of cuprate superconductors in the normal-state has been evaluated in terms of the full charge-spin recombination \cite{Feng16,Zhao16,Zhao17}, and can be expressed explicitly as,
\begin{eqnarray}\label{EGF}
G({\bf k},\omega)={1\over \omega-\varepsilon_{\bf k}-\Sigma_{1}({\bf k},\omega)},
\end{eqnarray}
where the bare electron excitation spectrum $\varepsilon_{\bf k}=-4t\gamma_{\bf k}+4t'\gamma'_{\bf k}+\mu$, with $\gamma_{\bf k}=({\rm cos}k_{\rm x}+ {\rm cos}k_{\rm y})/2$ and $\gamma'_{\bf k}={\rm cos}k_{\rm x}{\rm cos}k_{\rm y}$, while the electron self-energy $\Sigma_{1}({\bf k},\omega)$ arises from the interaction between electrons by the exchange of spin excitations, and has been given in Ref. \onlinecite{Feng16}. The electron function $A({\bf k},\omega)$ on the other hand can be obtained directly from the electron Green's function (\ref{EGF}) as,
\begin{eqnarray}\label{ESF}
A({\bf k},\omega)={2|{\rm Im}\Sigma_{1}({\bf k},\omega)|\over [\omega-\varepsilon_{\bf k}-{\rm Re}\Sigma_{1}({\bf k},\omega)]^{2}+[{\rm Im} \Sigma_{1}({\bf k},\omega)]^{2}},
\end{eqnarray}
where ${\rm Im}\Sigma_{1}({\bf k},\omega)$ and ${\rm Re}\Sigma_{1}({\bf k},\omega)$ are the corresponding imaginary and real parts of $\Sigma_{1}({\bf k},\omega)$, respectively.

\begin{figure}[h!]
\centering
\includegraphics[scale=0.5]{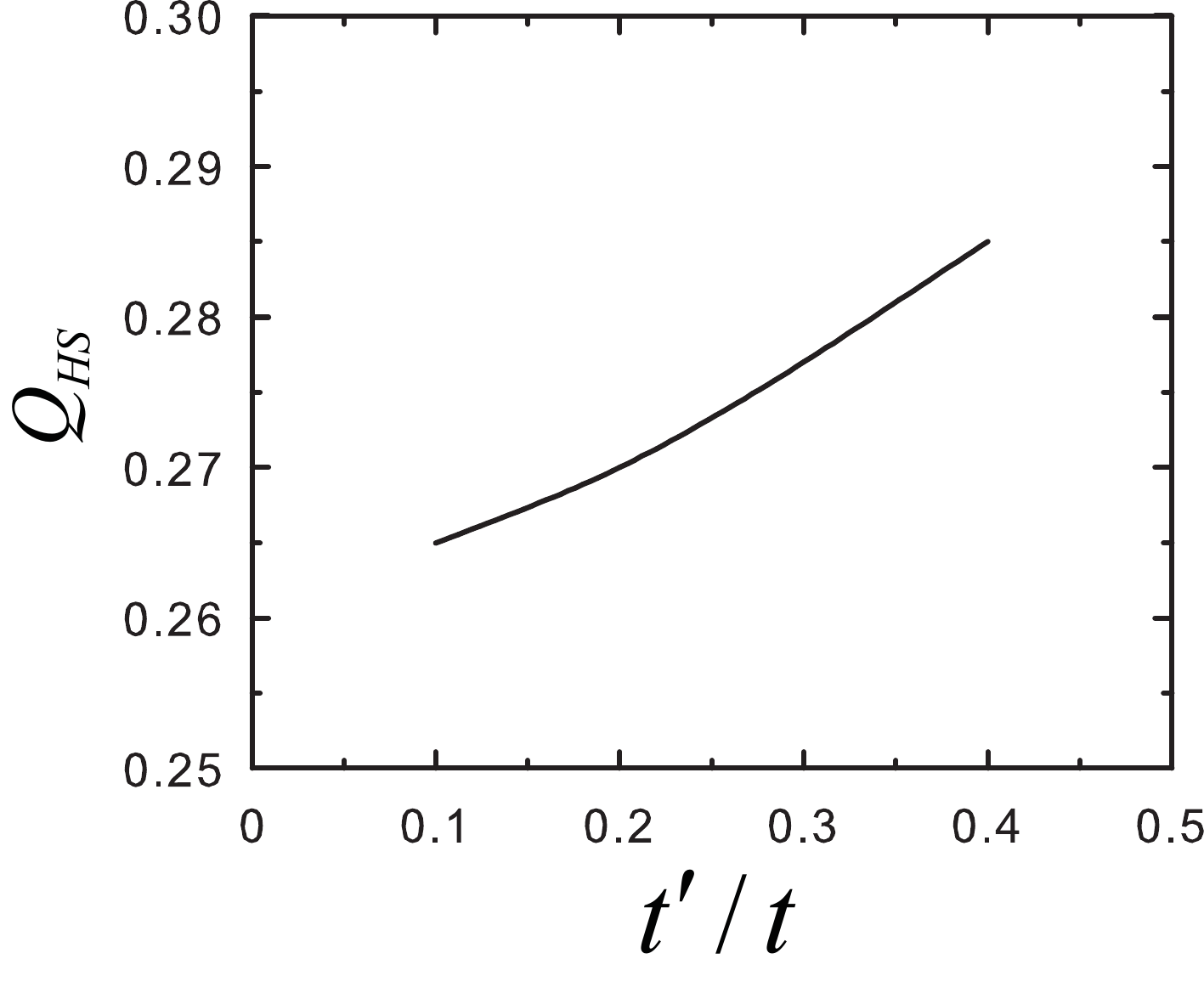}
\caption{The charge-order wave vector as a function of the second-neighbor hopping at $\delta=0.12$ for $t/J=2.5$ with $T=0.002J$.  \label{CD-t}}
\end{figure}

Now we turn to discuss the correlation between the charge-order wave vector $Q_{\rm CD}$ and second-neighbor hopping $t'$ in cuprate superconductors. As we have shown in our previous study \cite{Mou17,Feng16,Zhao16}, the charge-order state in cuprate superconductors is driven by the EFS instability, with the charge-order wave vector corresponding to the hot spots of the Fermi arcs. However, for the discussions of the correlation between the charge-order wave vector and second-neighbor hopping $t'$, we have performed a calculation for the electron spectral function $A({\bf k},0)$ in Eq. (\ref{ESF}) at zero energy with different values of $t'$, and the results of the maps of the spectral intensity $A({\bf k}, 0)$ at the Fermi energy and doping $\delta=0.12$ with temperature $T=0.002J$ for parameters (a) $t/J=2.5$ and $t'/t=0.1$, (b) $t/J=2.5$ and $t'/t=0.2$, (c) $t/J=2.5$ and $t'/t=0.3$, and (d) $t/J=2.5$ and $t'/t=0.4$ are plotted in Fig. \ref{spectrum}, where the highest peak heights on the contours are located at the hot spots, which appear always at the off-node places, in agreement with the ARPES experimental data \cite{Comin14,Shi08,Sassa11}. The results in Fig. \ref{spectrum} show clearly that there are two continuous contours in momentum space, which are labeled as ${\bf k}_{\rm F}$ and ${\bf k}_{\rm BS}$, respectively. However, the low-energy spectral weight at the contours ${\bf k}_{\rm F}$ and ${\bf k}_{\rm BS}$ around the antinodal region has been suppressed, and then the low-energy electron excitations occupy disconnected segments located at the contours ${\bf k}_{\rm F}$ and ${\bf k} _{\rm BS}$ around the nodal region. In particular, the tips of the disconnected segments on the contours ${\bf k}_{\rm F}$ and ${\bf k}_{\rm BS}$ converge at the hot spots to form a closed Fermi pocket, leading to a coexistence of the Fermi arc and Fermi pocket \cite{Zhao17,Yang08,Chang08,Meng09,Yang11}, where the disconnected segment at the first contour ${\bf k}_{\rm F}$ is so-called the Fermi arc, and is also defined as the front side of the Fermi pocket, while the other at the second contour ${\bf k}_{\rm BS}$ is associated with the back side of the Fermi pocket. On the other hand, the curvatures of the Fermi arc ${\bf k}_{\rm F}$ and back side of the Fermi pocket ${\bf k}_{\rm BS}$ are strongly sensitive to the second-neighbor hopping $t'$, and then the area of the Fermi pockets increases with the increase of $t'$, indicating that $t'$ plays a crucial role in the EFS reconstruction. Moreover, the most of the electron quasiparticles are accommodated at the hot spots, and then the electron quasiparticles are scattered between two hot spot regions connected by the charge-order wave vector $Q_{\rm HS}$ and the same electron quasiparticle scattering causes the charge-order formation \cite{Comin15,Comin14,Wu11,Chang12,Ghiringhelli12,Neto14,Campi15,Comin15a,Hashimoto15,Mou17,Feng16,Zhao16}. Our present results in Fig. \ref{spectrum} also show that the magnitude of the charge-order wave vector $Q_{\rm HS}$ increases with the increase of $t'$. To show this point more clearly, the result for the extracted charge-order wave vector $Q_{\rm HS}$ as a function of $t'$ at $\delta=0.12$ for $t/J=2.5$ with $T=0.002J$ is plotted in Fig. \ref{CD-t}, where $Q_{\rm HS}$ grows linearly with the increase of $t'$. It is thus shown that the experimentally observed differences of the magnitudes of the charge-order wave vector $Q_{\rm CD}$ in the different families of cuprate superconductors at the same doping concentration are attributed to the different values of $t'$. In particular, the charge-order wave vector connecting the hot spots is found to be $Q_{\rm HS}\sim 0.265$ in Fig. \ref{spectrum} for $t/J=2.5$ and $t'/t=0.1$, closely matching the experimental value \cite{Comin14} of the charge-order wave vector $Q_{\rm CD}\sim 0.26$ found in Bi$_{2}$Sr$_{2-x}$La$_{x}$CuO$_{6+\delta}$ at the doping regime around $\delta\approx 0.12$.

\begin{figure}[h!]
\centering
\includegraphics[scale=0.4]{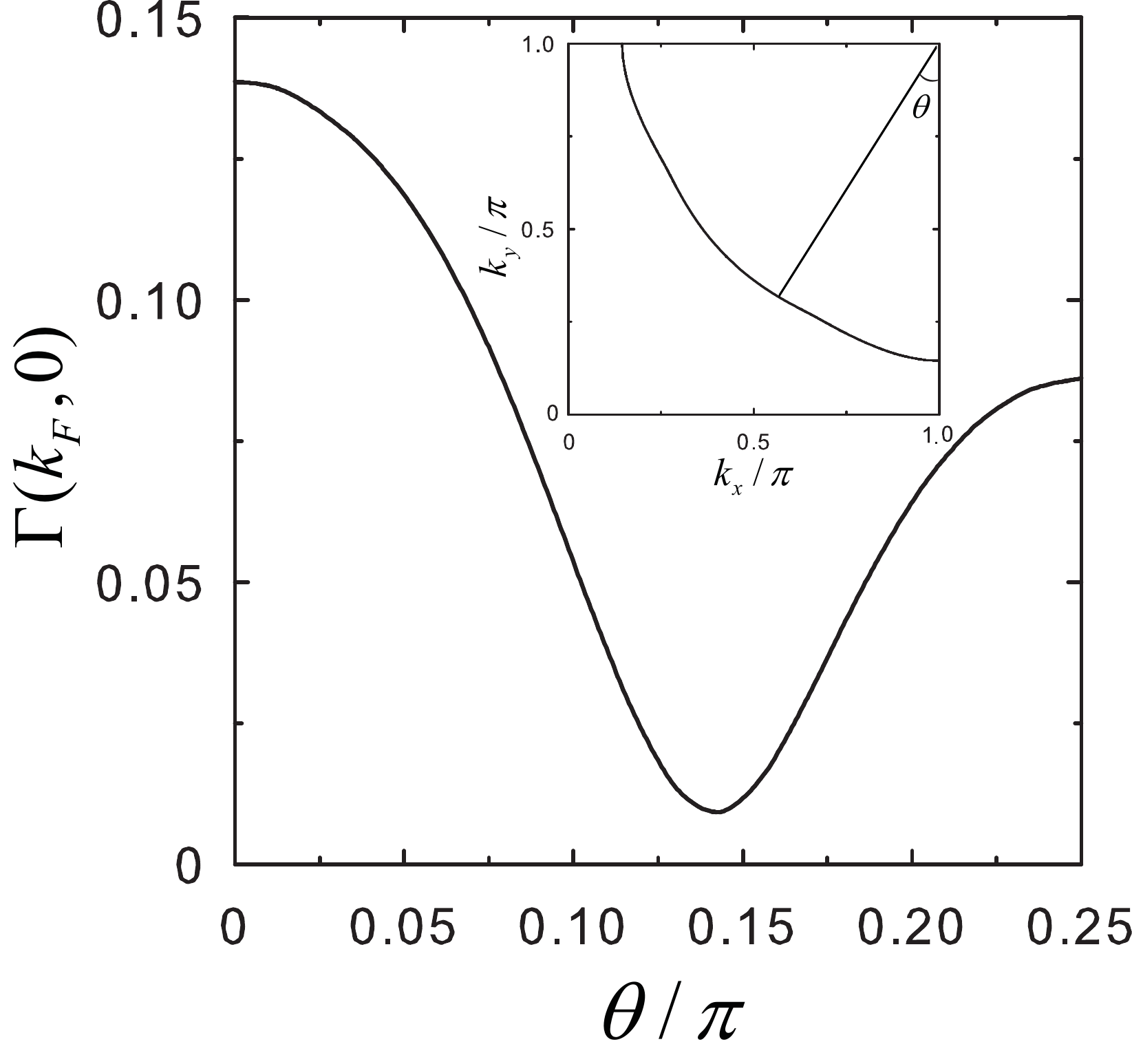}
\caption{The angular dependence of the electron quasiparticle scattering rate on the electron Fermi surface at $\delta=0.12$ with $t/J=2.5$, $t'/t=0.3$, $T=0.002J$. \label{scattering-rate}}
\end{figure}

The essential physics of the correlation between the charge-order wave vector $Q_{\rm CD}$ and second-neighbor hopping $t'$ in cuprate superconductors is the same as that of the charge-order state driven by the EFS instability \cite{Mou17,Feng16,Zhao16}. The locations of the continuous contours in momentum space are determined directly by the poles of the electron Green's function (\ref{EGF}) at zero energy, while the intensity of the low-energy electron excitation spectrum at the continuous contours is inversely proportional to the imaginary part of the electron self-energy ${\rm Im}\Sigma_{1}({\bf k},\omega)$. In particular, the dynamical electron quasiparticle scattering rate $\Gamma({\bf k},\omega)$ is directly related to this imaginary part of the electron self-energy as $\Gamma({\bf k},\omega)={\rm Im}\Sigma_{1}({\bf k},\omega)$. On the other hand, the electron self-energy $\Sigma_{1}({\bf k},\omega)$ in Eq. (\ref{EGF}) can be also rewritten as \cite{Feng16,Zhao16,Zhao17},
\begin{eqnarray}\label{CD-gap}
\Sigma_{1}({\bf k},\omega)\approx {[\bar{\Delta}_{\rm PG}({\bf k})]^{2}\over\omega+\varepsilon_{0{\bf k}}},
\end{eqnarray}
where the corresponding energy spectrum $\varepsilon_{0{\bf k}}$ and the momentum dependence of the pseudogap $\bar{\Delta}_{\rm PG}({\bf k})$ can be obtained directly from the electron self-energy $\Sigma_{1}({\bf k},\omega)$ and its antisymmetric part $\Sigma_{\rm 1o}({\bf k},\omega)$ as $\varepsilon_{0{\bf k}}=-\Sigma_{1}({\bf k},0)/\Sigma_{\rm 1o}({\bf k},0)$ and $\bar{\Delta}_{\rm PG}({\bf k})=\Sigma_{1}({\bf k},0)/ \sqrt{-\Sigma_{\rm 1o}({\bf k},0)}$, respectively. In this case, the dynamical electron quasiparticle scattering rate $\Gamma({\bf k},\omega)$ can be expressed in terms of the pseudogap as $\Gamma({\bf k},\omega)=2\pi[\bar{\Delta}_{\rm PG}({\bf k})]^{2}\delta(\omega+\varepsilon_{0{\bf k}})$. As a consequence of the presence of the pseudogap \cite{Zhao17}, the electron energy band has been split into the antibonding band and bonding band, respectively, and then the first contour ${\bf k}_{\rm F}$ in Fig. \ref{spectrum}, represents the contour in momentum space, where the electron antibonding dispersion along ${\bf k}_{\rm F}$ is equal to zero, while the second contour ${\bf k}_{\rm BS}$ in Fig. \ref{spectrum}, is the contour in momentum space, where the electron bonding dispersion along ${\bf k}_{\rm BS}$ is equal to zero. However, the pseudogap $\bar{\Delta}_{\rm PG}({\bf k})$ is strong dependence of momentum \cite{Feng16,Zhao16,Zhao17,Jing17}. To see this strongly anisotropic pseudogap in momentum space clearly, we plot the angular dependence of the electron quasiparticle scattering rate $\Gamma({\bf k}_{\rm F},0)$ on EFS at $\delta=0.12$ with $T=0.002J$ for $t/J=2.5$ and $t'/t=0.3$ in Fig. \ref{scattering-rate}. Moreover, we have also found that the angular dependence of the electron quasiparticle scattering rate $\Gamma({\bf k}_{\rm BS},0)$ along the back side of the Fermi pocket is very similar to that of $\Gamma({\bf k}_{\rm F},0)$. In Fig. \ref{scattering-rate}, it is shown clearly that $\Gamma({\bf k}_{\rm F},0)$ (then the pseudogap) has a strong angular dependence with the actual maximum at the antinode, which leads to that the low-energy electron spectral weight at the contours ${\bf k}_{\rm F}$ and ${\bf k}_{\rm BS}$ around the antinodal region is gapped out by the pseudogap. However, the actual minimum does not appear around the node, but locates exactly at the hot spot ${\bf k}_{\rm HS}$, which leads to that the tips of these disconnected segments on ${\bf k}_{\rm F}$ and ${\bf k}_{\rm BS}$ converge on the hot spots to form the closed Fermi pocket around the nodal region, generating a coexistence of the Fermi arcs and Fermi pockets \cite{Zhao17}. This special structure of the momentum dependent electron quasiparticle scattering rate in Fig. \ref{scattering-rate} also leads to that the remarkable peak-dip-hump structure in the electron spectrum of cuprate superconductors \cite{Dessau91,Ding96,Saini97,Fedorov99} is absent from the hot-spot directions \cite{Zhao16,Zhao17}. Furthermore, these electron quasiparticles are scattered between two hot spot regions connected by the characteristic wave vector $Q_{\rm HS}$, and then this electron quasiparticle scattering therefore induces the charge-order formation \cite{Feng16,Zhao16}. In particular, as shown in Fig. \ref{spectrum}, with the increase of the values of the second-neighbor hopping $t'$, the position of the hot spots moves towards to the nodes, which therefore leads to that the charge-order wave vector $Q_{\rm HS}$ connected by the hot spots on the straight Fermi arcs increases with the increase of $t'$. This is why the differences of the magnitudes of the charge-order wave vector $Q_{\rm CD}$ can be observed experimentally in the different families of cuprate superconductors at the same doping concentration.

In summary, we have discussed the correlation between the charge-order wave vector $Q_{\rm CD}$ and second-neighbor hopping $t'$ in cuprate superconductors based on the $t$-$t'$-$J$ model. Our results show that the magnitude of the charge-order wave vector $Q_{\rm CD}$ increases with the increase of $t'$, and then the experimentally observed differences of the magnitudes of the charge-order wave vector $Q_{\rm CD}$ among the different families of cuprate superconductors at the same doping concentration are attributed to the different values of $t'$.

\section*{Acknowledgements}

The authors would like to thank Dr. Deheng Gao and Professor Yongjun Wang for helpful discussions. HZ is supported by the National Natural Science Foundation of China (NSFC) under Grant Nos.11547034, and YM, DG and SF are supported by the National Key Research and Development Program of China under Grant No. 2016YFA0300304, and NSFC under Grant Nos. 11574032 and 11734002.

\end{document}